\documentclass[aps]{revtex4} 
\usepackage{txfonts}

\begin{document}
%------------------------------------------------------------------------------%

%------------------------------------------------------------------------------%
\title{Stability of Closed Timelike Geodesics.}
%------------------------------------------------------------------------------%

 \author{Val\'eria M. Rosa\footnote{e-mail: vmrosa@ufv.br}
  }

 \affiliation{
Departamento de Matem\'atica, Universidade Federal de Vi\c{c}osa, 
36570-000 Vi\c{c}osa, M.G., Brazil
}

\author{ Patricio  S. Letelier\footnote{e-mail: letelier@ime.unicamp.br}
 } 
 
\affiliation{
 Departamento de Matem\'atica Aplicada-IMECC,
Universidade Estadual de Campinas,
13081-970 Campinas,  S.P., Brazil}

\begin{abstract}
  The existence  and stability under linear perturbations 
of closed timelike geodesics (CTG) in Bonnor-Ward spacetime is 
studied in some detail. Regions where the CTGs exist and  are
 linearly stable are exhibited.
\end{abstract}

%\begin{keyword}
%Closed Timelike Geodesics \sep Linear Stability \sep Time Machines

%\PACS 04.20.Gz, 04.20.Dg, 04.20.Jb
%\end{keyword}

%\end{frontmatter}

\maketitle
In 1949 G\"odel found a solution to the Einstein field equation
with nonzero cosmological constant that admits closed timelike curves
(CTCs)~\cite{goedel}. It could be argued that the G\"odel solution is
without physical significance, since it corresponds to a rotating,
stationary cosmology, whereas the actual universe is expanding and
apparently non rotating. The van Stockum solution~\cite{stockum}, that
also contain CTCs, is physically inadmissible since refers to an
infinitely long cylinder. 

The existence of CTCs contradicts the usual notion of causality. Beyond
 the usual paradoxes, it seems to induce physical
impossibilities, like the necessity to work with negative energy
densities. One could speculate that these impossibilities will be
eliminated by quantum-gravitational effects.  All our experience 
seems to indicate that the physical laws do not allow  the
appearance of CTCs. This is that, essentially,  says the Chronology
Protection Conjecture (CPC) proposed by Hawking in
1992~\cite{hawking2}.

In some cases these CTCs can be disregarded 
because to have them one
ought to have an external force acting along the whole CTC, process that will
 consume a great  amount of energy. The energy needed to travel a CTC
 in G\"odel universe is computed in \cite{pfarr}.
 For geodesics this is not the case since  the  external force is null,
 therefore the considerations of energy does not apply in this case and 
we potentially  have a bigger  problem of breakdown of causality.

There exist examples of solutions of
vacuum Einstein's equations which contains CTCs that can represent the
exterior of physically admissible sources~\cite{bonnor}~\cite{bonnor2}.
In~\cite{bonnor} it is described the case of a massless spinning rod of
finite length. In~\cite{bonnor2} it is analyzed the CTCs in 
Kerr-Newman spacetime and in a solution of the Einstein equations for
 a source named Perjeon, due
to Perj\'es~\cite{perjes}, which represents a single charged,
rotating, magnetic object. This solution was also studied
independently by Israel and Wilson~\cite{israel}, and it is referred as
a PIW spacetime. In these three cases one expect that
the CTC region be covered by the source. The same does not happen when
we work with two Perjeons~\cite{bonnor2}. This last solution also has 
closed timelike geodesics (CTGs), one of these CTGs was exhibited 
in~\cite{bonnor2}.

The possibility that a spacetime
associated to a realistic model of matter may contain CTGs leads us to
ask how permanent is the existence of these curves. Perhaps, one may
rule out the CTGs by simple considerations about their linear
stability. Otherwise, if these curves are stable under linear
perturbations the conceptual problem associated  to their existence is enhanced.

%------------------------------------------------------------------------------%

The PIW metric is given by
\begin{equation}
ds^2=-f^{-1}h_{mn}dx^m dx^n+f(\omega_m dx^m+dt)^2,
\label{metric}
\end{equation}
where the three dimensional positive definite tensor $h_{mn}$ has zero
Ricci tensor and it will be taken as the usual three dimensional Euclidean
 metric in cylindrical coordinates, the electromagnetic field is given in 
terms of two
scalar potentials:
\begin{equation}
F_{4n}=\Phi_{,n},\; F^{ab}=\eta^{abm}f\psi_{,m},
\end{equation}
$\eta^{abm}$ being the Levi-Civita tensor related to 
$h_{mn},$ and $( ~)_{,n}=\partial/ \partial x^n$. The
entire solution is generated by two functions $L$ and $M$ that are
harmonic with respect to $h_{mn}$ by means of the  equations, 
\begin{eqnarray}
 & f = \frac{1}{4L^2+4M^2}, & \nonumber\\
 & \psi=\frac{1}{2}\epsilon\frac{M}{L^2+M^2}, & \nonumber\\
 & \Phi=-\frac{1}{2}\epsilon\frac{L}{L^2+M^2}, & \nonumber\\
 & \omega_{a,b}-\omega_{b,a}=8(ML_{,c}-LM_{,c})\eta_{abm}h^{mc}. & 
\end{eqnarray}
The Bonnor-Ward (BW) solution~\cite{bonnor3} refers to two Perjeons,  
with masses  $m_1$ and $m_2$,
placed on the $z$-axis at $z=\pm a,\,(a>0)$, with magnetic
moments ($\mu_1$ and $\mu_2$) also parallel to the $z$-axis, and
\begin{eqnarray}
&&L= (1+m_1/r_1+m_2/r_2)/2\nonumber\\
&&M = (\mu_1(z-a)/r_1^3+\mu_2(z+a)/r_2^3)/2\nonumber\\
&&r_1=\sqrt{\rho^2+(z-a)^2}\nonumber\\
&&r_2=\sqrt{\rho^2+(z+a)^2}\nonumber\\
&&\omega_a= -\rho^2\Omega\, \delta_{a}^\varphi.
\end{eqnarray}

We shall consider the particular case of BW solution~\cite{bonnor2},
\begin{eqnarray}
&&\Omega=\frac{\mu_1}{r_1^3}\Big(2+\frac{m_1}{r_1}\Big)+\frac{m_1\mu_2}{\rho^2}\Big(\frac{r_1}{2a^2r_2}+\frac{(z+a)(\rho^2+z^2-a^2)}{ar_1r_2^3}\Big)\nonumber\\
&&+\frac{\mu_2}{r_2^3}\Big(2+\frac{m_2}{r_2}\Big)+\frac{m_2\mu_1}{\rho^2}\Big(\frac{r_2}{2a^2r_1}-\frac{(z-a)(\rho^2+z^2-a^2)}{ar_2r_1^3}\Big)\nonumber\\
&&+\frac{\alpha}{\rho^2},
\label{omega}
\end{eqnarray}
with $\alpha$ restricted to the values
$\alpha=\pm(m_1\mu_2+m_2\mu_1)/(2a^2) (\ne 0)$ only.  When
$\alpha=(m_1\mu_2+m_2\mu_1)/(2a^2)$ the spacetime has a torsion line 
 singularity (TLS)
on $\rho=0$ for $z^2>a^2$ and in the other case the singularity
appears between the sources. In both cases there is no strut
singularity~\cite{bonnor2}. The presence of the TLSs is unavoidable in this 
case, they keep the two stationary spinning, charged, magnetic sources rotating 
in equilibrium (for torsions lines see~\cite{letelier1}\cite{letelier2}). In 
other words we have a solution with a TSL that extend to infinite and a 
 solution with a compact source. This last solution  is asymptotically flat. Even though the
first one does not represent a physically acceptable solution we shall
 consider it  for further analysis.

Let  us denote by $\gamma$  a  closed curve in BW spacetime 
given in its parametric form by, 
\begin{equation}
  t = t_0, \hspace{2cm}
  \rho = {\rm constant}, \hspace{2cm}
  \varphi \in [0,2\pi], \hspace{2cm} z=z_0,
  \label{CTC}
\end{equation}
where $t_0$, $\rho$ and $z_0$ are constants. The closed curve is
timelike when $g_{\varphi\varphi} > 0$, i.e.,
\begin{equation}
\Omega^2f^2 \rho^2-1>0.
\label{tl_cond}
\end{equation}
   The  four-acceleration of $\gamma$ is
\begin{eqnarray}
  \label{acceleration}
&& a^t=0, \\
&& a^{\rho}=\frac{\partial_{\rho}f\Omega^2\rho^3f^2+2f^3\Omega\rho^3\partial_{\rho}\Omega+4f^3\Omega^2\rho^2-2f+\rho\partial_{\rho}f}{2\rho(f^2\Omega^2\rho^2-1)}\dot{\varphi}^2, \\
&& a^{\varphi}=0, \\
&& a^z=\frac{\partial_zf\Omega^2\rho^2f^2+2f^3\Omega\rho^2\partial_z\Omega+\partial_zf}{2(f^2\Omega^2\rho^2-1)}\dot{\varphi}^2.
\end{eqnarray}
We shall restrict us to the special case,
\begin{equation}
 m_1=m_2=m, \; \mu_1=\mu_2=\mu.
\label{geo_cond_z}
\end{equation}
Furthermore, we take $z_0=0$ to have $a^z=0$.
The condition  $a^{\rho}=0$ gives us to different cases, depending on the
position of the TLS.

%------------------------------------------------------------------%
First we analise the case $\alpha
=-(m_1\mu_2+m_2\mu_1)/(2a^2)$   (TSL  between the particles). By assuming 
the restriction  (\ref{geo_cond_z})  and doing  $a^{\rho}=0$ we obtain the 
following relation for
$\mu$, $a$, $\rho$ and $m$,
\begin{eqnarray}
&&\mu=-1/4\Big(-2\big(4\rho^4m^3+8\sqrt{\rho^2+a^2}m^2\rho^4+2\sqrt{\rho^2+a^2}\rho^6-16a^2\rho^2m^3\nonumber\\
&&-19a^4m\rho^2+6\rho^6m-2a^2\sqrt{\rho^2+a^2}\rho^4+7a^4m^3-15a^2m\rho^4+2a^6m\nonumber\\
&&-32a^2\sqrt{\rho^2+a^2}m^2\rho^2+8a^4\sqrt{\rho^2+a^2}m^2-4a^4\sqrt{\rho^2+a^2}\rho^2)(a^6\sqrt{\rho^2+a^2}\nonumber\\
&&+10a^6m+3a^4\sqrt{\rho^2+a^2}\rho^2+28a^4m\rho^2+80a^4m^3+40a^4\sqrt{\rho^2+a^2}m^2\nonumber\\
&&+80a^2\sqrt{\rho^2+a^2}m^4+64a^2\sqrt{\rho^2+a^2}m^2\rho^2+3a^2\sqrt{\rho^2+a^2}\rho^4+32a^2m^5\nonumber\\
&&+112a^2\rho^2m^3+\sqrt{\rho^2+a^2}\rho^6+24\sqrt{\rho^2+a^2}m^2\rho^4+8\rho^6m+32\rho^4m^3\nonumber\\
&&+26a^2m\rho^4+16\sqrt{\rho^2+a^2}m^4\rho^2\big)\Big)^{1/2}(\rho^2+a^2)/\Big(4\rho^4m^3+8\sqrt{\rho^2+a^2}m^2\rho^4\nonumber\\
&&+2\sqrt{\rho^2+a^2}\rho^6-16a^2\rho^2m^3-19a^4m\rho^2+6\rho^6m-2a^2\sqrt{\rho^2+a^2}\rho^4\nonumber\\
&&+7a^4m^3-15a^2m\rho^4-32a^2\sqrt{\rho^2+a^2}m^2\rho^2+2a^6m+8a^4\sqrt{\rho^2+a^2}m^2\nonumber\\
&&-4a^4\sqrt{\rho^2+a^2}\rho^2\Big).
\label{geo_cond_rho1}
\end{eqnarray}
Using this form for $\mu$ and doing $\rho=n \,a$  we obtain 
for $g_{\varphi\varphi},$
\begin{eqnarray}
&&g_{\varphi\varphi}=-\big[2n^2(n^2+1)(2n^2-1)a^2+m\sqrt{n^2+1}(6n^4-11n^2-1)a 
\nonumber\\
&&+2m^2(2n^4-5n^2-1)\big]/\big[2n^2\big((n^2+1)(n^2-2)a^2+2m\sqrt{n^2+1}(n^2-4)a \nonumber\\
&&+m^2(2n^2-7)\big)\big].
\end{eqnarray}
We can find $n$,  the distance $a$, and the  mass $m$ such that $g_{\varphi\varphi}>0$.  We have closed timelike
geodesics for $n \in [0.708,\sqrt{2}]$ and $a>A\,m$,  for $n \in
(\sqrt{2},1.6385]$ and $a \in [A\,m,B\,m]$, and  for $n \in (1.6385,1.8872]$
and $a \in [0,B\,m]$, where
\begin{eqnarray}
&&A=\frac{\sqrt{1+6n^2+61n^4+60n^6-28n^8}-(6n^4-11n^2-1)}{4\sqrt{n^2+1}(2n^2-1)}, \nonumber\\
&&B=\frac{4-n^2+\sqrt{2+3n^2-n^4}}{\sqrt{n^2+1}(n^2-2)}.\nonumber
\end{eqnarray}
In the particular case $a=\rho$, the condition (\ref{tl_cond}) is
written as $(2\alpha-2\sqrt{2}m)/(2\alpha+5\sqrt{2}m)>0$ and if
$a>\sqrt{2}m$ the closed geodesic is timelike. In~\cite{bonnor2}
Bonnor and Steadman describe a CTG for $a=\rho=1/\sqrt{2}$, in this case the
closed curve $\gamma$ is a geodesic when (\ref{geo_cond_rho1}) reduces
to $\mu=(2m+1)^2/(\sqrt{10m^2+22m+4})$. The condition (\ref{tl_cond})
is written as $(1-4m)/(1+5m)>0$, so $\gamma$ is timelike when $m<1/4$.

%-------------------------------------------------------------------------%

Now we analise the case $\alpha =(m_1\mu_2+m_2\mu_1)/(2a^2)$ (two semi-infinity TLS). Assuming
that (\ref{geo_cond_z}) is true and making $a^{\rho}=0$ we obtain the
following relation for $\mu$, $a$, $\rho$ and $m$,
\begin{eqnarray}
&&\mu=\Big(\big(4a^8\sqrt{\rho^2+a^2}+2a^6\sqrt{\rho^2+a^2}\rho^2-2a^4\sqrt{\rho^2+a^2}\rho^4+\rho^6m^3\nonumber\\
&&+48a^6m^2\sqrt{\rho^2+a^2}+24a^8m-a^2\rho^6m-7a^4\rho^4m+18a^6\rho^2m+32a^6m^3\nonumber\\
&&+4a^2\rho^4m^3\big)\big(10a^6m+a^6\sqrt{\rho^2+a^2}+3a^4\sqrt{\rho^2+a^2}\rho^2+80a^4m^3+28a^4m\rho^2\nonumber\\
&&+40a^4\sqrt{\rho^2+a^2}m^2+3a^2\sqrt{\rho^2+a^2}\rho^4+32a^2m^5+26a^2m\rho^4+112a^2\rho^2m^3\nonumber\\
&&+64a^2\sqrt{\rho^2+a^2}m^2\rho^2+80a^2\sqrt{\rho^2+a^2}m^4+24\sqrt{\rho^2+a^2}m^2\rho^4+8\rho^6m\nonumber\\
&&+8a^4\rho^2m^3+\sqrt{\rho^2+a^2}\rho^6+16\sqrt{\rho^2+a^2}m^4\rho^2+32\rho^4m^3\big)\Big)^{1/2}(\rho^2+a^2)a^2/\nonumber\\
&&\big(4\sqrt{2}(4a^8\sqrt{\rho^2+a^2}+2a^6\sqrt{\rho^2+a^2}\rho^2-2a^4\sqrt{\rho^2+a^2}\rho^4+\rho^6m^3\nonumber\\
&&+48a^6m^2\sqrt{\rho^2+a^2}+24a^8m-a^2\rho^6m-7a^4\rho^4m+18a^6\rho^2m\nonumber\\
&&+32a^6m^3\rho^2m^3+8a^4+4a^2\rho^4m^3)\rho\big)
\label{geo_cond_rho2}
\end{eqnarray}
Using this form for $\mu$ and $\rho=n\,a,$ as before, we get
\begin{eqnarray}
&&g_{\varphi\varphi}=\big[-2m^2(n^4-n^2+4)+m^2\sqrt{n^2+1}(n^4+9n^2-8)a+2(2n^4+n^2-1)a^2\big],\nonumber\\
&&/\big[2\big(m^2(n^4+8)-2m\sqrt{n^2+1}(n^2-4)a-(n^2+1)(n^2-1)a^2\big)\big].
\end{eqnarray}
Also in this case, we can find $n$, a distance
$a,$ and a mass $m$ such that $g_{\varphi\varphi}>0$. In particular, we find $a$ and $m$ such that
$g_{\varphi\varphi}>0$ for $n>\sqrt{2}$ and $C\,m<a<D\,m$, where
\begin{eqnarray}
&&C=\frac{n^2\sqrt{n^4+50n^2+17}-n^2-9n+8}{2\sqrt{n^2+1}(2n^2-1)},\nonumber\\
&&D=\frac{n^2-4-n^2\sqrt{n^2-1}}{\sqrt{n^2+1}(2-n^2)}.\nonumber
\end{eqnarray}
For example, we have CTGs for $n=1.5$ and $0.3482\,m<a<9.4644\,m$,  for
$n=2$ and $0.2725\,m<a<1.5491\,m$,  for $n=5$ and $0.2456\,m<a<0.8652\,m$, and
for $n=10$ and $0.1704\,m<a<0.9127\,m$.

%\vspace{5mm}
%------------------------------------------------------------------------------%

A generic CTC $\gamma$ satisfies the system of equations given by 
\begin{equation}
\frac{D}{ds} \dot{X}^{\mu}= F^\mu(X),
\label{CTCsystem}
\end{equation}
where $\frac{D\  W^{\alpha}}{ds} $ is the covariant derivative of
the vector field $W^{\alpha}$ along $\gamma(s)$ and $F^\mu$ is an external force
per unit of mass.

Let $\tilde{\gamma}$ be the curve obtained from $\gamma$ after a
perturbation ${\bf \xi}$,
i.e., $\tilde{X}^{\mu}=X^{\mu}+\xi^{\mu}$. 
The system of differential equation
satisfied by the perturbation ${\bf \xi}$ is~\cite{shirokov}~\cite{letelier3},
\begin{equation}
\frac{d^2\xi^{\alpha}}{ds}+2\Gamma^{\alpha}_{\beta \mu}\delta
u^{\beta}u^{\mu}+\Gamma^{\alpha}_{\beta
\mu,\lambda}\xi^{\lambda}u^{\beta}u^{\mu}=F^{\alpha}_{,\lambda}\xi^{\lambda}.
\label{systemPerturbation}
\end{equation}

For the closed curve (\ref{CTC}) the
system~(\ref{systemPerturbation}), for case where (\ref{geo_cond_z})
holds, $z=z_0=0$,  and $\mu$, $a$, $\rho$, and $m$ are related in  such a way that
$a^{\rho}=0$ ($F^{\alpha}=0$), reduces to
\begin{equation}
\begin{array}{l}
\ddot{\xi}^0+k_1\dot{\xi}^1=0,\\[5mm] 
\ddot{\xi}^1+k_2\dot{\xi}^0+k_3\xi^1=0,\\[5mm]
\ddot{\xi}^2+k_4\dot{\xi}^1=0,\\ [5mm]
\ddot{\xi}^3+k_5\xi^3=0,
\end{array}
\label{sistemaCTG}
\end{equation}
where $k_1=2\Gamma^0_{21}\dot{\varphi}$, $k_2=2\Gamma^1_{20}\dot{\varphi}$,
 $k_3=\Gamma^1_{22,1}\dot{\varphi}^2$, $k_4=2\Gamma^2_{21}\dot{\varphi}$,
 $k_5=\Gamma^3_{22,3}\dot{\varphi}^2$. The condition for $\gamma$ to be
 timelike, when it is parametrized by the proper time $s$, is $
 \dot{X}^{\mu} \dot{X}_{\mu} = 1, $ where the overdot indicates
 derivation with respect to $s$. For the curve $\gamma$ this last
 condition gives us
\begin{equation}
\dot{\varphi}^2=\frac{f}{\rho^2(f^2\rho^2w^2-1)}.
\end{equation}

The  solution of (\ref{sistemaCTG})  is
\begin{equation}
\begin{array}{l}
\xi^0=-k_1(c_3\sin(\varpi s+c_4)/\varpi+\lambda s)+c_1\,s+c_5,\\ 
\xi^1=c_3\cos(\varpi s+c_4)+\lambda, \\
\xi^2=-k_4(c_3\sin(\varpi s+c_4)/\varpi+\lambda s)+c_2\,s+c_6,\\ 
\xi^3=c_7\cos(\sqrt{k_5} s+c_8),
\end{array}\label{exactpert}
\end{equation}
where $c_i,\;i=1,\dots,8$ are integration constants, $\varpi=\sqrt{k_3-k_1k_2}$, 
and $\lambda = -k_2c_1/\varpi^2$. The explicit form of $\varpi$ and $k_5$ depending 
on the distance $a$, the mass $m$ and the parameter $n$ are cumbersome; they will be 
presented elsewhere.

To describe the
regions where the CTGs are linearly stable, first, we  fix $n$ ($\rho=n\, a$) and 
then we find  $a$ and $m$ such that
$\varpi^2>0$ and $k_5>0$.
%------------------------------------------------------------------------------%
When the TLS is between the sources, the closed curve
$\gamma$, satisfying (\ref{geo_cond_z}),  $z=z_0=0$,  and (\ref{geo_cond_rho1}) is a
CTG linearly stable when $n \in [1.5,1.8]$. For example, we have
$\varpi^2>0$ and $k_5>0$ for $n=1.5$ and $7.4788\,m<a<7.988\,m$, and 
for $n=1.8$ and $0.1046\,m<a<0.5279\,m$.

For the case $a=\rho=1/\sqrt{2}$ given by Bonnor and Steadman~\cite{bonnor2}, we have
\begin{equation}
\frac{\varpi^2}{k_5}=-2\frac{(84m^4+235/2m^2+7/2+35m+183m^3)(m+2)}{(1+2m)^2(116m+33/2+53m^3+170m^2)}
\end{equation}
that is always negative. In this case the CTG is not linearly
stable. This is a particular case of $a=\rho$, where
\begin{equation}
\frac{\varpi^2}{k_5}=\frac{(14\alpha^4+70\sqrt{2}m\alpha^3+235m^2\alpha^2+183\sqrt{2}m^3\alpha+84m^4)(4\alpha+m\sqrt{2})}{-(\alpha+m\sqrt{2})^2(66\alpha^3+232\sqrt{2}m\alpha^2+340m^2\alpha+53m^3\sqrt{2})}
\end{equation}

%------------------------------------------------------------------------------%

When we have two semi-infinite torsion TLS the closed curve
$\gamma$, satisfing (\ref{geo_cond_z}), $z=z_0=0$,  and (\ref{geo_cond_rho2}), is a
CTG linearly stable when $n \ge 1.5$. For example, we have
$\varpi^2>0$ and $k_5>0$ for $n=1.5$ and $8.7913\,m<a<9.3079\,m$, for
$n=2$ and $0.4353\,m<a<1.2147\,m$, for $n=5$ and
$0.2904\,m<a<0.4576\,m$, and for $n=10$ and $0.2743\,m<a<0.5014\,m$.

%------------------------------------------------------------------------------%

In summary, we found that there are linearly stable closed
timelike geodesics in BW spacetime. There are two cases depending
whether the torsion line singularity is between the sources or not. 
We note that the existence of CTGs or their
linear stability does not depend on the value of the mass parameter $m$.

It is interesting to see that in one case the  CTGs does 
not enclosed the TLS. 
One can think that the TLS is a mathematical representation 
of a physical device to keep the particles rotating in such a way that allow 
the existence  of CTGs. In particular, the infinite TLSs may be replaced by
some physical structure with motors. The amount of energy to keep this
 configurations stable is not known and its computation is under study. In 
this case the very notion of energy is not clear.

The analysis performed in this letter is the simplest and should be considered 
as a first step to understand the CTGs stability. To regard this example as a 
counter example of the CPC from a pure classical approach one need  to consider
 second order perturbations. Moreover one  would like to have a kind of structural 
stability of the CTGs, e.g., that the change of the spacetime by the addition of
 a small amount of  matter does not destroy the CTGs. But, if they are  second 
order and structural stable we will have an evident challenge to the General 
Relativity Theory or our usual notion of causality will need a deep revision.

V.M.R.  thanks Departamento de Matem\'atica-UFV for giving the
conditions to finish this work which was partially supported  by
PICDT-UFV/CAPES. PSL thanks the partial financial
support of FAPESP and CNPq.

%------------------------------------------------------------------------------%

%------------------------------------------------------------------------------%
\end{document}